\documentclass[seceq,preprint]{ptptex}
\usepackage{amsmath,amssymb,mathtools}
\newcommand{\mspd}[1]{[#1]_G}
\newcommand{\epsi}{(\eta\Psi)}

\newcommand{\qpsi}{(Q_G\Psi)}

\newcommand{\qxi}{(Q_G\Xi)}

\notypesetlogo                       %comment in if to eliminate PTPTeX 
\preprintnumber[3cm]{%<-- [..]: optional width of preprint # column.
YITP-14-52}
\title{%        %You can use \\ for explicit line-break.
First-Order Equations of Motion\\ for Heterotic String Field Theory
}

\author{%       %Use \scshape for the family name.
Hiroshi \textsc{Kunitomo}\footnote{%
E-mail:\  {\tt kunitomo@yukawa.kyoto-u.ac.jp}}
}

\inst{%     %Affiliation, neglected when [addenda] or [errata].
Yukawa Institute for Theoretical Physics, Kyoto University, \\
Kyoto 606-8502, Japan
}

\abst{%         %This abstract is neglected when [addenda] or 
We consider the equations of motion of the full heterotic string 
field theory including both the Neveu-Schwarz and the Ramond sectors.
It is shown that they can be formulated in the form of an infinite 
number of first-order equations for an infinite number of independent 
string fields. We prove that the conventional equations of motion are 
obtaned by solving the extra equations for the extra string fields
with a certain assumptions at the linearized level. The conventional 
gauge transformations are also obtained from those in this first-order 
formulation, which is clarified by deriving some lower oder 
transformations explicitly. 
}
\begin{document}

\maketitle

\section{Introduction}

The Wess-Zumino-Witten(WZW)-like formulation is one of the promising approaches to 
construct consistent superstring field theories, which was first proposed for 
an open superstring field theory,\cite{Berkovits:1995ab} and afterwards
extended to a heterotic string field theory.\cite{Okawa:2004ii,Berkovits:2004xh}
In this formulation, string fields are defined using the large Hilbert space 
which includes the zero mode of the worldsheet fermion $\xi$ coming from 
bosonization of the superghost $(\beta,\gamma)$.\cite{Friedan:1985ge}
A major advantage of this formulation is that the picture number conservation is 
built in without explicitly introducing the picture-changing operator, whose collision 
causes a breakdown of gauge invariance. The action of the Neveu-Schwarz (NS) sector is
written as a WZW action using a pure-gauge string field, which is an analog
of the Maurer-Cartan form, $g^{-1}dg$.

In spite of this success of the NS sector, the Ramond (R) sector
in the WZW-like formulation has been less studied so far. This might be 
because one of the recent motivation to study string field theories is 
to find analytic solutions, including that for tachyon condensation, 
which does not require the R sector. Another reason is the difficulty
to construct a covariant action of the R sector 
consistently with the picture number conservation.\cite{Berkovits:2001im}
There is no question, however, that 
the full string field theory including the R sector has to be constructed
not only for theoretical consistency but also 
for studying various problems related to the supersymmetry,
such as the perturbative finiteness and the supersymmetry breaking.
Therefore, as the second best option, the (gauge-invariant) equations of motion
was given for the full open superstring\cite{Berkovits:2001im} 
as in the case of a self-dual $(2n+1)$ form in $4n+2$ dimensions.

From this reason, we attempted to similarly construct the equations of motion of 
the full heterotic string field theory,\cite{Kunitomo:2013mqa}
which are nonpolynomial not only in the NS string field $V$ but also in the R string
field $\Psi$. Expanding in powers of $\Psi$,
we found explicit forms of equations of motion and gauge transformations
up to some lower order in $\Psi$, or to all orders in $\Psi$ for certain categories 
of terms. In addition, we were also able to find that a gauge invariance requires 
the complete equations of motion to be in the form of\footnote{
The composite fields $B_{-1/2}$ and $B_{-1}$ were denoted as $\Omega$ and $\Sigma$
in Ref.~\citen{Kunitomo:2013mqa}. The phase convention of the R string field $\Psi$ 
is also changed, which causes the sign difference in front of the second 
term of (\ref{eomNS}).}
\begin{subequations}\label{eom}
 \begin{align}
  \eta G(V)+\frac{\kappa}{2}\mspd{(B_{-\frac{1}{2}})^2}+Q_GB_{-1} =&0,\label{eomNS}\\
  Q_GB_{-\frac{1}{2}} =& 0,\label{eomR}
 \end{align}
\end{subequations}
under the assumption that the NS string field $V$ couples to the R string 
field $\Psi$ only through the pure gauge string field $G(V)$ in the shifted BRST 
operator $Q_G$ and the shifted string products $\mspd{\cdots}$ (G-ansatz).
Here $B_{-1/2}$ and $B_{-1}$ are the sum of the terms with the picture numbers 
$P=-1/2$ and $-1$, respectively, constructed using the shifted string products 
of $\Psi$, $\eta\Psi$ and $Q_G\Psi$. Their explicit forms were obtained 
at the lower orders in $\Psi$\cite{Kunitomo:2013mqa} 
and in general expected to be determined by requiring a consistency condition 
of (\ref{eomNS}),
\begin{equation}
 \eta\left(\eta G+\frac{\kappa}{2}\mspd{(B_{-\frac{1}{2}})^2}+Q_GB_{-1}\right) = 0,
\label{condition}
\end{equation}
although we have not succeeded in proving that the condition has 
a unique nontrivial solution. The purpose of this paper to establish a method 
to give an explicit expression of the equations of motion and gauge transformations 
of the full heterotic string field theory up to an arbitrary order in $\Psi$. 
The condition (\ref{condition}) will be replaced with an endless sequence of
consistency equations including an infinite number of composite fields $B_{-n/2}$
$(n\ge1)$ to be determined.
These infinite number of equations can also be written in a single equation with 
the same form as the equation of motion of the closed \text{bosonic} string field theory. 
We can reinterpret them as the first-order equations of motion 
for the infinite number of independent fields $B_{-n/2}$
under a certain assumption at the linearized level. 
We will prove that the conventional equations of motion (\ref{eom}) are 
obtained by solving the extra equations in the first-order formulation, 
which provides a desired method to give an explicit form of (\ref{eom}).
An infinite number of gauge transformations in the first-order formulation
can also be written in the form of the single transformation with the same 
form as that in the closed bosonic string field theory. 
These symmetries are used to fix an ambiguity 
in the solutions of the extra equations, and reduced
to the conventional gauge symmetries.

In \S\ref{OSSFT} of this paper, we will briefly summarize the known results on
the equations of motion and the gauge transformations of the full open superstring 
field theory. This will be useful for examining those of the heterotic string 
field theory in \S\ref{HSFT}. 
After introducing some fundamental building blocks and summarizing 
their properties, the equations of motion, with an endless sequence of 
the consistency conditions, will be written in a single equation including 
an infinite number of composite string fields to be determined.
We will prove that this equation can be interpreted as the first-order equations 
of motion, and provides a procedure to obtain the conventional equations of motion 
(\ref{eom}) in a specific form. The conventional gauge transformations can also be 
obtained from those in the first-order formulation, which will be clarified 
by deriving some lower order transformations explicitly. 
Section \ref{condis} is devoted to conclusion and discussion.
An appendix is added to give some concrete results on the equations of motion and
the gauge transformations.

\newpage

\section{Equations of motion and gauge transformations for open superstring}\label{OSSFT}

In this section, we summarize the known results on the open superstring field 
theory,\cite{Berkovits:2001im} focusing on the equations of motion and the gauge 
transformations.
In the WZW-type open superstring field theory, 
the NS string field $\Phi$ and the R string field $\Psi$
are both Grassmann even, and carry the ghost and picture numbers $(G,P)=(0,0)$ 
and $(0,1/2)$, respectively. The equations of motion of the full theory, 
including both the NS and the R sectors, are given by
\begin{subequations}\label{openEOM}
\begin{align}
\eta J(\Phi)+(\eta\Psi)^2=&0,\label{eomsecondNS}\\
Q_J\eta\Psi=&0,\label{eomsecondR}
\end{align}
\end{subequations}
where $J(\Phi)=A_Q=e^{-\Phi}(Qe^\Phi)$ is a pure-gauge solution
of the equation of motion of the open \textit{bosonic} string field theory,
namely, it identically satisfies
\begin{equation}
QJ(\Phi)+J(\Phi)^2=0.\label{eJ}
\end{equation}
The BRST operator $Q_J$ shifted by $J(\Phi)$ is defined on a general open 
superstring field $A$ by
\begin{equation}
Q_JA=QA+J(\Phi)A-(-1)^{|A|}AJ(\Phi),
\end{equation}
and is nilpotent $(Q_J)^2=0$ due to (\ref{eJ}).
The equations of motion (\ref{openEOM}) have the symmetry under
the gauge transformations,
\begin{subequations}\label{open gtf}
\begin{align}
A_\delta=& Q_J\Lambda_0+\eta\Lambda_1
-\epsi\Lambda_{\frac{1}{2}}-\Lambda_{\frac{1}{2}}\epsi,\\
\delta\Psi=&Q_J\Lambda_{\frac{1}{2}}+\eta\Lambda_{\frac{3}{2}}+\Psi(\eta\Lambda_1)-(\eta\Lambda_1)\Psi,
\end{align}
\end{subequations}
where $A_\delta=e^{-\Phi}(\delta e^\Phi)$
and the gauge parameters $\Lambda_{n/2}$, with $n=$ even (odd), 
are Grassmann odd NS (R) string fields carrying $(G,P)=(0,n/2)$.

The equations of motion (\ref{openEOM}) are equivalent to
the first-order equations,
\begin{subequations}\label{oeom}
\begin{align}
&QJ+J^2 = 0,\label{oeom1}\\
&Q_J\Omega = 0,\label{oeom2}\\
&\eta J+\Omega^2 = 0,\label{oeom3}\\
&\eta\Omega = 0,\label{oeom4}
\end{align}
\end{subequations}
where the NS string field $J$ and the R string field $\Omega$ are 
the \textit{independent} Grassmann odd string fields,
and carry the ghost and picture numbers $(G,P)=(1,0)$ and $(1,-1/2)$,
respectively. 
The conventional (second-order) equations of motion (\ref{openEOM}) are obtained 
by solving (\ref{oeom1}) and (\ref{oeom4}) 
using unconstrained string fields $\Phi$ and $\Psi$ as
\begin{equation}
 J=J(\Phi),\qquad \Omega=\eta\Psi,\label{potential}
\end{equation}
and then substitute them into (\ref{oeom3}) and (\ref{oeom2}).
This first-order equations of motion (\ref{oeom}) can also be written
in a single equation with the same form as the equation of motion of 
the open bosonic string field theory,
\begin{equation}
 \hat{Q}\hat{A}+\hat{A}^2=0,\label{openEOM1st}
\end{equation}
where
\begin{equation}
 \hat{Q}=Q+\eta,\qquad \hat{A}=J+\Omega.
\end{equation}
Four equations (\ref{oeom}) are obtained by expanding (\ref{openEOM1st})
into the different pictures, $P=0,-1/2,-1$ and $-3/2$, each of
which has to vanish separately.

The gauge-transformations can also be written 
in a single transformation with the same form as that
of the open bosonic string field theory,
\begin{equation}
\delta\hat{A}=\hat{Q}\hat{\sigma}+\hat{A}\hat{\sigma}-\hat{\sigma}\hat{A},\label{opengt}
\end{equation}
with $\hat{\sigma}=\sigma_{1/2}+\sigma_0$.
These gauge parameters $\sigma_n$ are, however, not free but constrained
because (\ref{opengt}) is equivalent to the gauge transformations of $J$ and $\Omega$
with two constraints,
\begin{subequations}\label{ogauge}
\begin{align}
0=& Q_J\sigma_{\frac{1}{2}},\label{ogauge1}\\
\delta J =& Q_J\sigma_0
+\Omega\sigma_{\frac{1}{2}}-\sigma_{\frac{1}{2}}\Omega,\label{ogauge2}\\
\delta \Omega =& \eta\sigma_{\frac{1}{2}}+
\Omega\sigma_0-\sigma_0\Omega,\label{ogauge3}\\
0 =& \eta\sigma_0,\label{ogauge4}
\end{align}
\end{subequations}
obtained by expanding both sides of (\ref{opengt}) into different pictures.
The first and the last equations, (\ref{ogauge1}) and (\ref{ogauge4}), impose,
under (\ref{oeom1}), that
\begin{equation}
\sigma_{\frac{1}{2}}=Q_J\Lambda_{\frac{1}{2}},
\qquad 
\sigma_0=\eta\Lambda_1.
\end{equation}
Substituting this into (\ref{ogauge2}) and (\ref{ogauge3}), we obtain
the gauge transformations
\begin{subequations}\label{first-order gtf}
\begin{align}
 \delta J =& Q_J\eta\Lambda_1+\Omega (Q_J\Lambda_{\frac{1}{2}})-(Q_J\Lambda_{\frac{1}{2}})\Omega,\\
 \delta \Omega =& \eta Q_J\Lambda_{\frac{1}{2}}+\Omega(\eta\Lambda_1)-(\eta\Lambda_1)\Omega,
\end{align}
 \end{subequations}
which generate the symmetry of (\ref{oeom}). 
The conventional gauge transformation (\ref{open gtf}) can be read 
from (\ref{first-order gtf})
using an analog of a Maurer-Cartan (MC) equation,
\begin{equation}
\delta J(\Phi)=Q_J(e^{-\Phi}\delta e^\Phi).\label{MCEopen}
\end{equation}
The additional symmetries generated by $\Lambda_0$ and $\Lambda_{\frac{3}{2}}$ 
come from the degree of freedom keeping $J(\Phi)$ and $\eta\Psi$ invariant, respectively.

\section{Equations of motion and gauge transformations for heterotic string}\label{HSFT}

Now let us consider the equations of motion and the gauge transformations for
the heterotic string field theory. The NS string field $V$ and the R string field
$\Psi$ in the heterotic string field theory are both Grassmann odd,
and carry the ghost and the picture numbers $(G,P)=(1,0)$ and $(1,1/2)$,
respectively.
They satisfy the subsidiary conditions for closed string fields.\cite{Okawa:2004ii}

\subsection{String products and their identities}\label{sp and id}

At the beginning, we briefly summarize basic ingredients of the heterotic string
field theory.\cite{Okawa:2004ii,Berkovits:2004xh} 
The algebraic structure of the heterotic string field theory includes a BRST operator $Q$
and graded-commutative multi-linear string products, $[B_1,\cdots,B_n]$ with $n\ge2$.
The fundamental identities that the string products satisfy are
\begin{align}
 0 =& Q[B_1,\cdots,B_n]+\sum^n_{i=1}(-1)^{(B_1+\cdots+B_{i-1})}[B_1,\cdots,QB_i,\cdots,B_n]\nonumber\\
&
+\sum_{\underset{l+k=n}{\{i_l,j_k\}}}\sigma(i_l,j_k)[B_{i_1},\cdots,B_{i_l},[B_{j_1},\cdots,B_{j_k}]],
\label{main}
\end{align}
where the factor $\sigma(i_l,j_k)$ is defined to be the sign picked up when one
rearranges the sequence $\{Q,B_1,\cdots,B_n\}$ into the order 
$\{B_{i_1},\cdots,B_{i_l},Q,B_{j_1},\cdots,B_{j_k}\}$. The concrete realization of the
string products can be constructed based on the polyhedral 
overlapping conditions.\cite{Nonpolynomial,Kugo:1989aa,Kugo:1989tk}
In addition, there are two important operators $X=\eta$ and $\delta$ 
to construct equations of motion and gauge transformations. 
They act on the string products as derivations; that is,
\begin{equation}
 X[B_1,\cdots,B_n]=(-1)^X\sum^n_{i=1}(-1)^{X(B_1+\cdots+B_{i-1})}[B_1,\cdots,XB_i,\cdots,B_n].\label{derivation}
\end{equation}
The pure-gauge closed string field $B_Q=G(V)$ is an analog of $J(\Phi)=e^{-\Phi}(Qe^\Phi)$ in the open
superstring field theory, and defined as a pure-gauge solution of 
the equation of motion of the closed \textit{bosonic} string field theory:\cite{Berkovits:2004xh}
\begin{equation}
 QG(V)+\sum_{n=2}^\infty\frac{\kappa^{n-1}}{n!}[G(V)^n] = 0.\label{1st eq}
\end{equation}
The BRST operator and string products shifted by $G(V)$ can be defined by
\begin{align}
 Q_G B\equiv QB+\sum_{m=1}^\infty\frac{\kappa^m}{m!}[G^m,B],\\
[B_1,\cdots,B_n]_G\equiv \sum_{m=0}^\infty\frac{\kappa^m}{m!}[G^m,B_1,\cdots,B_n],
\end{align}
for general string fields $\{B,B_1,\cdots,B_n\}$, and play an important role below. 
Owing to (\ref{1st eq}), this shifted BRST operator $Q_G$ is nilpotent and acts on the shifted 
products $\mspd{\cdots}$ in exactly the same way as (\ref{main}).\cite{Zwiebach:1992ie}  
In contrast, the operator $X$ is neither graded commutative with $Q_G$ nor derivation
on the shifted products, but satisfies
\begin{subequations} 
\begin{align}
&Q_G(XB)-(-1)^XX(Q_GB)=-\kappa\mspd{XG,B},\label{relation1}\\
&X\mspd{B_1,\cdots,B_n}=\sum_{i=1}^n(-1)^{X(1+B_1+\cdots+B_n)}\mspd{B_1,\cdots,XB_i,\cdots,B_n}
\nonumber\\
&\hspace{3.3cm}
+(-1)^X\kappa\mspd{XG,B_1,\cdots,B_n}.\label{relation2}
\end{align}
 \end{subequations}
The MC equation for the heterotic string is given by
\begin{equation}
 XG(V)=Q_GB_X(V),\label{MCEhet}
\end{equation}
where $B_X(V)$ is an analogous string field to $A_X=e^{-\Phi}(Xe^\Phi)$ 
in the open superstring field theory,\cite{Berkovits:2004xh} whose explicit form is not
necessary in this paper. If we take $X=\eta$, this leads to a nontrivial identity,
\begin{equation}
 Q_G(\eta G(V))\equiv 0,\label{id1}
\end{equation}
which is essential to derive the general form (\ref{eom}) of the equations of motion.

\subsection{Equations of motion and their consistency}\label{eom con foe}

Using the pure-gauge closed string field $G(V)$,
the equation of motion for the NS sector of the heterotic string is written
as\cite{Berkovits:2004xh}
\begin{equation}
 \eta G(V)=0.\label{NSsectorEOM}
\end{equation}
This has the symmetry under the gauge transformations,
\begin{equation}
 B_\delta(V) = Q_G\Lambda_0+\eta\Lambda_1,\label{NSsectorgauge}
\end{equation}
since
\begin{equation}
\delta(\eta G(V))=-\kappa\mspd{\eta G(V),\eta\Lambda_1}.
\end{equation}
In particular, the equation of motion (\ref{NSsectorEOM}) is \textit{invariant}
under the $\Lambda_0$-transformation since it keeps $G(V)$ invariant.
When we incorporate the coupling to the R sector, we assume that the NS string 
field $V$ appears in the correction terms only through $G(V)$ 
in the BRST operator $Q_G$ and the string products $\mspd{\cdots}$ and, 
in consequence, the full equations of motion are as well invariant
under the $\Lambda_0$-gauge transformation:
\begin{equation}
 B_\delta(V) = Q_G\Lambda_0,\qquad \delta\Psi=0.
\end{equation} 
Under this assumption, which we call G-ansatz, 
the identity (\ref{id1}) restricts the form of the full equations
of motion as (\ref{eom}).
In this paper, we impose not only (\ref{condition}) but also
the consistency of the R equation of motion to determine the explicit forms 
of $B_{-1/2}$ and $B_{-1}$:
\begin{subequations}\label{con}
\begin{align}
 \eta\left(\eta G+\frac{\kappa}{2}\mspd{(B_{-\frac{1}{2}})^2}+Q_GB_{-1}\right) =& 0,
\label{con1}\\
\eta Q_GB_{-\frac{1}{2}} =& 0.\label{con2}
\end{align}
\end{subequations}
After some calculation, one can show that these conditions 
(\ref{con}) are equivalent, up to the equations of motion (\ref{eom}), to
\begin{subequations}\label{oseq}
\begin{align}
&
\eta B_{-\frac{1}{2}} +\kappa\mspd{B_{-\frac{1}{2}},B_{-1}}
+\frac{\kappa^2}{3!}\mspd{(B_{-\frac{1}{2}})^3} 
+ Q_GB_{-\frac{3}{2}} = 0,\label{con omega} \\
&\eta B_{-1} +\kappa\mspd{B_{-\frac{1}{2}},B_{-\frac{3}{2}}} 
+\frac{\kappa}{2}\mspd{(B_{-1})^2}\nonumber\\
&\hspace{0.4cm}
+\frac{\kappa^2}{2}\mspd{(B_{-\frac{1}{2}})^2,B_{-1}} 
+\frac{\kappa^3}{4!}\mspd{(B_{-\frac{1}{2}})^4}+Q_GB_{-2} = 0,\label{con sigma}
\end{align}
\end{subequations}
with new composite fields, $B_{-3/2}$ and $B_{-2}$ to be determined.
We further have to impose the consistency conditions of (\ref{oseq})
to determine these new fields, which yield
\begin{subequations}\label{gpeq}
\begin{align}
&\eta B_{-\frac{3}{2}}+\kappa\mspd{B_{-\frac{1}{2}},B_{-2}}
+\kappa\mspd{B_{-1},B_{-\frac{3}{2}}}
+\frac{\kappa^2}{2}\mspd{(B_{-\frac{1}{2}})^2,B_{-\frac{3}{2}}}\nonumber\\
&\hspace{20mm}
+\frac{\kappa^2}{2}\mspd{B_{-\frac{1}{2}},(B_{-1})^2}
+\frac{\kappa^3}{3!}\mspd{(B_{-\frac{1}{2}})^3,B_{-1}}
+\frac{\kappa^4}{5!}\mspd{(B_{-\frac{1}{2}})^5}+Q_GB_{-\frac{5}{2}}=0,\\
%%%%%%%%%%%
&\eta B_{-2}+\kappa\mspd{B_{-\frac{1}{2}},B_{-\frac{5}{2}}}
+\kappa\mspd{B_{-1},B_{-2}}
+\frac{\kappa}{2}\mspd{(B_{-\frac{3}{2}})^2}\nonumber\\
&\hspace{3mm}
+\frac{\kappa^2}{2}\mspd{(B_{-\frac{1}{2}})^2,B_{-2}}
+\kappa^2\mspd{B_{-\frac{1}{2}},B_{-1},B_{-\frac{3}{2}}}
+\frac{\kappa^2}{3!}\mspd{(B_{-1})^3}
+\frac{\kappa^3}{3!}\mspd{(B_{-\frac{1}{2}})^3,B_{-\frac{3}{2}}}\nonumber\\
&\hspace{16mm}
+\frac{\kappa^3}{4}\mspd{(B_{-\frac{1}{2}})^2,(B_{-1})^2}
+\frac{\kappa^4}{4!}\mspd{(B_{-\frac{1}{2}})^4,B_{-1}}
+\frac{\kappa^5}{6!}\mspd{(B_{-\frac{1}{2}})^6}+Q_GB_{-3}=0,
\end{align}
\end{subequations}
up to (\ref{eom}) and (\ref{oseq}), with new composite fields $B_{-\frac{5}{2}}$ and $B_{-3}$. 
This sequence of consistency conditions does not terminate but produce an infinite 
number of equations with an infinite number of composite fields $B_{-n/2}$
to be determined. 
The resultant infinite number of equations, however, can simply be written in
a single equation with the same form as the equation of motion of 
the closed \textit{bosonic} string field theory, 
\begin{equation}
 \hat{Q}\hat{B}+\sum_{m=2}^\infty\frac{\kappa^{m-1}}{m!}[\hat{B}^m]=0,
\label{FOeom}
\end{equation}
where
\begin{equation}
 \hat{Q}=Q+\eta,\qquad \hat{B}=\sum_{n=0}^\infty B_{-\frac{n}{2}}.
\end{equation}
The composite fields $B_{-\frac{n}{2}}$ with $n=$ even (odd)
are Grassmann even NS (R) string fields, and have $(G,P)=(2,-n/2)$. 
The infinite number of equations are obtained by expanding (\ref{FOeom}) 
into different pictures which have to hold separately.
It is easy to confirm that the first seven of the equations actually
coincide with (\ref{1st eq}), (\ref{eomR}), (\ref{eomNS}), (\ref{oseq}) and (\ref{gpeq})
with the identification $B_0=G(V)$.
The consistency of an infinite number of equations 
can now be shown at once by
\begin{align}
 \hat{Q}\left(\hat{Q}\hat{B}+\sum_{n=2}^\infty\frac{\kappa^{n-1}}{n!}[\hat{B}^n]\right)=&
-\sum_{m=1}^\infty\frac{\kappa^m}{m!}[\hat{B}^m,\left(\hat{Q}\hat{B}
+\sum_{n=2}^\infty\frac{\kappa^{n-1}}{n!}[\hat{B}^n]\right)],\label{conB}
\end{align}
which is a consequence of the fact that $\hat{Q}$ gives the same algebraic structure as $Q$:
it satisfies the same relation as (\ref{main}). 
The transformation of composite fields $B_{-n/2}$ are also written 
in a single transformation with the same form as that of the closed bosonic string field theory:
\begin{equation}
 \delta\hat{B} = \hat{Q}\hat{\sigma} 
+ \sum_{m=1}^\infty \frac{\kappa^m}{m!}[\hat{B}^m, \hat{\sigma}],\label{Het gauge}
\end{equation}
with the composite parameter $\hat{\sigma} = \sum_{n=-1}^\infty \sigma_{-n/2}$,
the details of which will be explained later.

\subsection{First-order formulation}\label{proof} 

The equation (\ref{FOeom}) can also be interpreted as the first-order equation
for the infinite number of \textit{independent} string fields $B_{-n/2}$
with $(G,P)=(2,-n/2)$ equivalent to the equations of motion (\ref{eom}). 
In order to see this, let us expand it into different picture numbers, 
and examine the equation at each picture as the first-order equations of motion.

First of all, the equation with the largest picture, $P=0$, 
\begin{equation}
 QB_0+\sum_{n=2}^\infty\frac{\kappa^{n-1}}{n!}[(B_0)^n] = 0,
 \label{1st eq 2}
\end{equation}
can be solved using the unconstrained NS string field $V$ as
\begin{equation}
B_0=G(V).
\end{equation}
We substitute this solution into the remaining equations below,
and consider the　$V$ and $\tilde{B}=\sum_{n=1}^\infty B_{-n/2}$
as the independent string fields.  
The next two equations at $P=-1/2$ and $-1$ have the same form
as the equations of motion (\ref{eomR}) and (\ref{eomNS}), respectively:
\begin{subequations}\label{EOM}
 \begin{align}
&Q_GB_{-\frac{1}{2}} = 0,\label{eom2}\\
&\eta G+\frac{\kappa}{2}\mspd{(B_{-\frac{1}{2}})^2} + Q_GB_{-1} = 0.\label{eom3}
\end{align}
\end{subequations}
Here, however, the string fields $B_{-1/2}$ and $B_{-1}$ are two
independent string fields.
The remaining equations $P\le-3/2$, which we call the subsidiary equations hereafter, 
can be written as
\begin{equation}
 S_{-\frac{(n+2)}{2}}\ \equiv\ \eta B_{-\frac{n}{2}}+Q_GB_{-\frac{(n+2)}{2}}
+\sum_{m=2}^{n+2}\frac{\kappa^{m-1}}{m!}
\mspd{\left(\tilde{B}^m\right)_{-\frac{(n+2)}{2}}}=0,\qquad (n\ge1),
\label{subsidiary}
\end{equation}
where $(\cdots)_P$ denotes the projection onto the terms with the picture 
number $P$ in total. 
The notation $S_{-(n+2)/2}$ are introduced to denote the left hand side of 
the subsidiary equations, with the subscript indicating their picture number, 
for later use.
As with the open superstring case in \S\ref{OSSFT}, the conventional equations 
of motion are obtained from (\ref{EOM}) if we can determine $B_{-1/2}$ and 
$B_{-1}$ as the composite fields of the $V$ and $\Psi$ by solving
(a part of) the subsidiary equations. Let us next show that this is in fact the case
if we assume, as \textit{an initial condition}, that only
the $B_{-1/2}$ of $B_{-n/2}$ is nontrivial (nonzero) at the linearized level. 
Under this assumption, all the subsidiary equations become trivial
at the linearized order, except for
\begin{equation}
S_{-\frac{3}{2}}^{(1)} = \eta B_{-\frac{1}{2}}^{(1)}=0,\label{linearized eom4}
\end{equation} 
which can be solved using the unconstrained R string field $\Psi$ as
\begin{equation}
 B_{-\frac{1}{2}}^{(1)}=\eta\Psi.\label{linearized omega}
\end{equation}
We explicitly specified the order in $\Psi$ as a superscript. 
So far, the argument is completely parallel to that of the open superstring field theory. 
In the heterotic string case, however, the linearized solution (\ref{linearized omega}) 
receives higher order corrections, and also make the infinite number of 
$B_{-n/2}$ nontrivial. 
If we plug (\ref{linearized omega}) into the subsidiary equations
(\ref{subsidiary}), we have, at the $(n+2)$-th order,
\begin{align}
S_{-\frac{(n+2)}{2}}^{(n+2)} =& \eta B_{-\frac{n}{2}}^{(n+2)}
+\frac{\kappa^{n+1}}{(n+2)!}\mspd{(\eta\Psi)^{n+2}}\nonumber\\
=&\eta\left(B_{-\frac{n}{2}}^{(n+2)}
-\frac{\kappa^{n+1}}{(n+2)!}\mspd{\Psi,\epsi^{n+1}}\right)
+\frac{\kappa^{n+2}}{(n+2)!}\mspd{\Psi,\epsi^{n+1},\eta G}=0.\label{sube1-1} 
\end{align}
Under the equation of motion (\ref{eom3}), these equations can be solved by
\begin{equation}
 B_{-\frac{n}{2}}^{(n+2)} = 
 \frac{\kappa^{n+1}}{(n+2)!}\mspd{\Psi,\epsi^{n+1}}, 
\label{leading}
\end{equation}
except for the $\eta$-exact terms. It is not necessary to consider this ambiguity
because the equations of motion (\ref{FOeom}) in the first-order formulation
are invariant under the gauge transformation (\ref{Het gauge}) with 
the \textit{independent} parameters $\sigma_{-n/2}$ for $n\ge0$
(and $\sigma_{1/2}=Q_G\Lambda_{1/2}$ constrained by $P=1/2$ component
of (\ref{FOeom}): $0=Q_G\sigma_{1/2}$).
The ambiguity coming from the $\eta$-exact terms in $B_{-n/2}$ for $n\ge2$ 
can be removed using this symmetry since they have the form
\begin{equation}
 \delta B_{-\frac{n}{2}}=\eta\sigma_{-\frac{(n-2)}{2}}
+\cdots,
\qquad (n\ge2).
\label{FOgauge}
\end{equation}
The ambiguity in $B_{-1/2}$, on the other hand, cannot similarly be gauged away, 
but can be absorbed into the redefinition of $\Psi$ owing to its leading order 
form (\ref{linearized omega}). 

The subsidiary equations can similarly be solved under the equations of motion, 
and determine $B_{-n/2}$ order by order in $\Psi$, except for apparent
ambiguities coming from the $\eta$-exact terms which can be removed in the same manner.
General procedure is, however, not so simple because the subsidiary equations 
are consistent only under the equations of motion, which breaks the independence 
of the equations at different orders.
In order to clarify what the difficulty is, let us examine the consistency 
equations of the subsidiary equations (\ref{subsidiary}), 
\begin{align}
\eta S_{-\frac{(n+2)}{2}}=&-Q_GS_{-\frac{(n+4)}{2}}
-\sum_{l=1}^{n+1}\sum_{m=1}^{n-l+2}\frac{\kappa^m}{m!}
\mspd{\left(\tilde{B}^m\right)_{-\frac{(n-l+2)}{2}},S_{-\frac{(l+2)}{2}}}\nonumber\\
&-\sum_{m=1}^{n+3}\frac{\kappa^m}{m!}
\mspd{\left(\tilde{B}^m\right)_{-\frac{(n+3)}{2}},E_{-\frac{1}{2}}}
-\sum_{m=1}^{n+2}\frac{\kappa^m}{m!}
\mspd{\left(\tilde{B}^m\right)_{-\frac{(n+2)}{2}},E_{-1}},\label{conSub}
\end{align}
obtained from (\ref{conSub}) and written in the form suitable
for the order-by-order analysis. 
For the $S$ at the given order in the left hand side,
the $S$s in the right hand side are the lower orders in $\Psi$.
The left hand side of the equations of motion (\ref{EOM}) are 
simply denoted as $E_{-1/2}$ and $E_{-1}$, respectively: 
\begin{equation}
E_{-\frac{1}{2}}=Q_GB_{-\frac{1}{2}},\qquad
E_{-1}=\eta G+\frac{\kappa}{2}\mspd{(B_{-\frac{1}{2}})^2}+Q_GB_{-1}.
\end{equation}
Substituting the lowest order solution (\ref{linearized omega}) into (\ref{conSub}),
we have the consistency equations at the first nontrivial, $(n+2)$-th, order:
\begin{equation}
\eta S_{-\frac{(n+2)}{2}}^{(n+2)}=
-\frac{\kappa^{n+2}}{(n+2)!}\mspd{(\eta\Psi)^{n+2},\eta G},
\end{equation}
which guarantees, from the triviality of the $\eta$-cohomology, that
$S_{-(n+2)/2}^{(n+2)}$ can be written as the sum of 
the $\eta$-exact part and the term including $\eta G$ as (\ref{sube1-1}). 
The point we should note is that, owing to the equations of motion,
the consistency equations do not hold separately in each order,
which prevent to solve the subsidiary equations at different orders 
independently. We have to take into account the higher order terms 
neglected in solving (\ref{sube1-1}) to consider the subsidiary equations 
at higher orders.
The simplest prescription to avoid this cumbersome procedure
is to modify the subsidiary equations (\ref{subsidiary}) as
\begin{align}
S(1)_{-\frac{(n+2)}{2}} =& S_{-\frac{(n+2)}{2}}
-f(1)_{-\frac{(n+2)}{2}}(E_{-\frac{1}{2}},E_{-1})\nonumber\\
\equiv& S_{-\frac{(n+2)}{2}}
-\frac{\kappa^{n+2}}{(n+2)!}\mspd{\Psi,\epsi^{n+1},E_{-1}}=0, 
\label{subsidiary2}
\end{align}
keeping the equivalence to the original ones under the \textit{exact} equation 
of motion $E_{-1}=0$. These new equations give, at the $(n+2)$-th order,
\begin{equation}
S(1)_{-\frac{(n+2)}{2}}^{(n+2)}=\eta\left(B_{-\frac{n}{2}}^{(n+2)}
-\frac{\kappa^{n+1}}{(n+2)!}\mspd{\Psi,\epsi^{n+1}}\right)=0,\label{exact1}
\end{equation}
which are now solved by (\ref{leading}) independently of the equations of motion,
and therefore make no contribution to the higher orders.
All the higher order contributions mentioned above are now incorporated 
in the improvement term. If we further replace $B_{-n/2}^{(n+2)}$ 
with the solution of $(\ref{exact1})$, namely (\ref{leading}),
the modified subsidiary equations (\ref{subsidiary2}) at the next, $(n+4)$-th, 
order are again given by the sum of the $\eta$-exact part and the terms 
including either $Q_G\eta\Psi$ or $\eta G$. Thus we can further modify 
(\ref{subsidiary2}) so as to be $\eta$-exact at the $(n+4)$-th order, 
as computed in the Appendix explicitly.

Repeating this procedure, we can prove by mathematical induction that
the subsidiary equations can be solved for $B_{-n/2}$ order by order in $\Psi$.
Suppose that the subsidiary equations can be modified as
\begin{subequations}\label{induction assumption}
 \begin{align}
S(k)_{-\frac{(n+2)}{2}} =& S(k-1)_{-\frac{(n+2)}{2}}
-f(k)_{-\frac{(n+2)}{2}}(E_{-\frac{1}{2}},E_{-1})\nonumber\\
=&S_{-\frac{(n+2)}{2}}
-\sum_{i=1}^{k}f(i)_{-\frac{(n+2)}{2}}(E_{-\frac{1}{2}},E_{-1})=0,
\label{subeq}
\end{align}
so as to be $\eta$-exact at the $(n+2k)$-th order,
\begin{equation}
S(k)_{-\frac{(n+2)}{2}}^{(n+2k)} = 
\eta\left(B_{-\frac{n}{2}}^{(n+2k)}+\cdots\right),
\label{n2k}
\end{equation}
by replacing $B_{-n/2}^{(n+2i)}$ for $0\le i\le k-1$
with the solutions of the lower order equations,
\begin{equation}
S(i)_{-(n+2)/2}^{(n+2i)}=0,\qquad (0\le i\le k-1).\label{lower sub}
\end{equation}
\end{subequations}
The improvement terms represented by $f(i)_{-(n+2)/2}(E_{-1/2},E_{-1})$
are assumed to contain either $E_{-1/2}$ or $E_{-1}$ for the equivalence
to the original equations.
For these modified subsidiary equations, we can find the similar consistency 
equations to (\ref{conSub}) using those of the equations of motion 
given in (\ref{conEOM}).
Although it is complicated to compute them explicitly, 
we can see that they have the similar structure to (\ref{conSub}),
namely, $\eta S(k)$ can be written as the sum of the terms including $S(k)$ 
and those including either $E_{-1/2}$ or $E_{-1}$, as required by 
the consistency. It is also easy to see that all the $S(k)$s but $\eta S(k)$
are again lower order in $\Psi$ than $\eta S(k)$.
From this structure with the induction assumptions (\ref{induction assumption})
and the triviality of the $\eta$-cohomology, 
we can conclude that replacing the solutions of
\begin{equation}
S(k)_{(n+2)/2}^{n+2k}=0,\label{on sub} 
\end{equation}
with $B_{-n/2}^{(n+2k)}$ in $S(k)$, we have
\begin{equation}
S(k)_{-\frac{(n+2)}{2}}^{(n+2(k+1))}
=\eta\left(B_{-\frac{n}{2}}^{(n+2(k+1))}+\cdots\right)
+f(k+1)_{-\frac{(n+2)}{2}}(Q_G\eta\Psi,\eta G),
\end{equation}
where the terms represented by $f_{-(n+2)/2}(Q_G\eta\Psi,\eta G)$ 
contain either $Q_G\eta\Psi$ or $\eta G$. 
Thus, if we further modify the subsidiary equations as
\begin{equation}
S(k+1)_{-\frac{(n+2)}{2}}\equiv S(k)_{-\frac{(n+2)}{2}} 
-f(k+1)_{-\frac{(n+2)}{2}}(E_{-\frac{1}{2}},E_{-1})=0,\label{subeq2}
\end{equation}
the new subsidiary equations are $\eta$-exact at the $(n+2(k+1))$-th order:
\begin{equation}
S(k+1)_{-\frac{(n+2)}{2}}^{(n+2(k+1))} = 
\eta\left(B_{-\frac{n}{2}}^{(n+2(k+1))}+\cdots\right) = 0.
\end{equation}
Hence by the mathematical induction the subsidiary equations
can be modified so that we have
\begin{equation}
S(k)_{-\frac{(n+2)}{2}}^{(n+2k)} = 
\eta\left(B_{-\frac{n}{2}}^{(n+2k)}+\cdots\right) = 0,
\end{equation}
for an arbitrary $k\in\mathbb{N}$,
which can be solved for $B_{-n/2}^{(n+2k)}$ and 
determine $B_{-n/2}$ order by order in $\Psi$.
The proof is completed.

In this way, the first-order formulation 
gives a systematic prescription to give an explicit expression 
of the conventional (second order) equations of motion (\ref{eom}). 
It is worth mentioning that all the infinite number of 
$B_{-n/2}$ are needed at each order to determine the equations 
of motion at the next order.
We will further illustrate this prescription 
with some concrete lower order results in the Appendix.

\subsection{Gauge transformations}\label{gtf}

In the first-order formulation, there is the gauge invariance 
(\ref{Het gauge}) with the infinite number of independent parameters. 
These parameters, however, are not independent in the conventional 
formulation since we need to use these symmetries to fix the ambiguity 
in solving the subsidiary equations. 
They are restricted so as to be consistent with the specific forms 
of $B_{-n/2}$ $(n\ge1)$. In this section, we clarify how the conventional 
gauge transformations are obtained from (\ref{Het gauge})
by explicitly finding the gauge transformations 
of $V$ and $\Psi$ up to some lower order.

Let us examine the transformation (\ref{Het gauge}) at each picture. 
The argument is again parallel to that for the case of the open superstring
at the linearized level. 
The component of the largest picture number $P=1/2$ leads to the constraint
$\sigma_{1/2}=Q_G\Lambda_{1/2}$.
Since the symmetry generated by $\sigma_{1/2}$ is not necessary to fix
the the ambiguity in the solutions of the subsidiary equations, 
the parameter $\Lambda_{1/2}$  generates 
a gauge symmetry in the conventional formulation without any further restriction.
We substitute it into the following transformations, and only consider
the $\sigma_{-n/2}$ $(n\ge0)$ as the parameters to be determined.
The next component at $P=0$ gives the transformation law
\begin{align}
 \delta B_0=&Q_G\sigma_0+\kappa\mspd{B_{-\frac{1}{2}},Q_G\Lambda_{\frac{1}{2}}}\nonumber\\
=&Q_G\left(\sigma_0-\kappa\mspd{B_{-\frac{1}{2}},\Lambda_{\frac{1}{2}}}\right)
-\kappa\mspd{Q_GB_{-\frac{1}{2}},\Lambda_{\frac{1}{2}}}.
\end{align}
This is consistent with $B_0=G(V)$, up to the equation of motion (\ref{eom2}), 
and we obtain
\begin{equation}
 B_\delta(V)=\sigma_0-\kappa\mspd{B_{-\frac{1}{2}},\Lambda_{\frac{1}{2}}}+Q_G\Lambda_0.
\label{tfV}
\end{equation}
The last term comes from the freedom that keeps $G(V)$ invariant.
The parameter $\sigma_0$ is not independent but 
constrained as with the other parameters.
The remaining transformations at $P\le-1/2$ have to be studied order by order in $\Psi$.
At the linearized level, only the $\sigma_0$ can be non-zero, 
except for $\sigma_{1/2}=Q_G\Lambda_{1/2}$,
from the consistency with the assumption that only $B_{-1/2}$ 
is nontrivial. The only nontrivial relation in (\ref{Het gauge}) is therefore
\begin{equation}
 0=\eta\sigma^{(0)}_0,\label{eta gauge}
\end{equation}
at $P=-1$, which restricts $\sigma_0$ as $\sigma_0^{(0)}=\eta\Lambda_1$
at the leading order. As the consequence, the leading order transformation
of $B_{-1/2}^{(1)}=\eta\Psi$ can be obtained from (\ref{Het gauge}) as
\begin{align}
(\delta B_{-\frac{1}{2}})^{(1)}=&\delta^{(0)}\left(\eta\Psi\right)\nonumber\\
=&\eta\left(Q_G\Lambda_{\frac{1}{2}}-\kappa\mspd{\Psi,\eta\Lambda_1}\right)
-\kappa^2\mspd{\eta G,\Psi,\eta\Lambda_1},\label{pleading}
\end{align}
from which we can read the gauge transformation of $\Psi$ at the leading order:
\begin{equation}
 \delta^{(0)}\Psi=Q_G\Lambda_{\frac{1}{2}}-\kappa\mspd{\Psi,\eta\Lambda_1}
+\eta\Lambda_{\frac{3}{2}}.
\label{psitf0}
\end{equation}

The leading order parameter $\sigma_0^{(0)}$ makes the infinite number of 
parameters $\sigma_{-n/2}$ nontrivial at the $(n+2)$-th order 
through the relations at $P\le-1$ in (\ref{Het gauge}),
\begin{align}
 0=&\eta\sigma_{-\frac{n}{2}}^{(n+2)}
+\frac{\kappa^{n+2}}{(n+2)!}\mspd{(B_{-\frac{1}{2}}^{(1)})^{n+2},\sigma_0^{(0)}}
\qquad (n\ge0),\nonumber\\
=&\eta\left(\sigma_{-\frac{n}{2}}^{(n+2)}
-\frac{\kappa^{n+2}}{(n+2)!}\mspd{\Psi,\epsi^{n+1},\eta\Lambda_1}\right)
-\frac{\kappa^{n+2}}{(n+2)!}\mspd{\eta G,\Psi,\epsi^{n+1},\eta\Lambda_1}.
\end{align}
Thus the parameters at the $(n+2)$-th order are restricted as\footnote{
We again fixed the ambiguity coming from the $\eta$-exact terms
because it can also be gauged away using the symmetry of the gauge transformation (\ref{Het gauge}):
$\delta\hat{\sigma}=\hat{Q}\hat{\tau}+\sum_{m=1}^\infty\frac{\kappa^m}{m!}[\hat{B}^m,\hat{\tau}]$
with $\hat{\tau}=\sum_{n=-2}^\infty\tau_{-n/2}$. The details will be discussed elsewhere.}
\begin{equation}
 \sigma_{-\frac{n}{2}}^{(n+2)}=\frac{\kappa^{n+2}}{(n+2)!}\mspd{\Psi,\epsi^{n+1},\eta\Lambda_1},
\qquad (n\ge0).
\end{equation}
Then the next order transformation of $B_{-1/2}$ in (\ref{Het gauge}) leads to
\begin{align}
\eta\delta^{(2)}\Psi+\delta^{(0)}\left(\frac{\kappa^2}{3!}\mspd{\Psi,\epsi^2}\right)
=&\frac{\kappa^3}{2}\mspd{\mspd{\epsi^2},\Psi,\eta\Lambda_1}
+Q_G\left(\frac{\kappa^3}{3!}\mspd{\Psi,\epsi^2,\eta\Lambda_1}\right)\nonumber\\
&+\frac{\kappa^3}{2}\mspd{\eta\Psi,\mspd{\Psi,\eta\Psi,\eta\Lambda_1}}
+\frac{\kappa^3}{3!}\mspd{\mspd{\Psi,\epsi^2},\eta\Lambda_1}\nonumber\\
&+\frac{\kappa^2}{2}\mspd{\epsi^2,Q_G\Lambda_{1/2}},\label{pnextleading}
\end{align}
where the first term in the right hand side
comes from $\mathcal{O}(\Psi^3)$ contribution of the previous order
transformation (\ref{pleading})
through the equations of motion. 
After some calculation, the next order transformation $\delta^{(2)}\Psi$ can 
be read as
\begin{align}
 \delta^{(2)}\Psi=
&\frac{\kappa^3}{3!}\mspd{\Psi,Q_G\Psi,\eta\Psi,\eta\Lambda_1}
-\frac{\kappa^3}{3!}\mspd{\Psi,\mspd{\Psi,\eta\Psi,\eta\Lambda_1}}
+\frac{\kappa^3}{3}\mspd{\mspd{\Psi,\eta\Psi},\Psi,\eta\Lambda_1}\nonumber\\
&-\frac{\kappa^2}{3}\mspd{\Psi,\eta\Psi,Q_G\Lambda_{\frac{1}{2}}}
+\frac{\kappa^2}{3!}\mspd{\Psi,\eta\Psi,\eta\Lambda_{\frac{3}{2}}}.
\end{align}
The higher order gauge transformations, $B_\delta$ (or equivalently $\sigma_0$)
and $\delta\Psi$, can similarly be obtained order by order in $\Psi$ 
from the gauge transformation (\ref{Het gauge}).

\section{Conclusion and discussion}\label{condis}

In this paper, we have established a method to give an explicit expression 
of the equations of motion and the gauge transformations
for the full heterotic string field theory.
An infinite number of extra composite fields have been introduced
with an infinite number of consistency conditions.
All of these equations have been written in a single equation, 
which has the same form as the equation of motion of 
the closed bosonic string field theory.
We have shown that these new equations can be interpreted as 
the first-order equations of motion under an assumption at the linearized level.
Then the conventional equations of motion have been obtained 
by solving extra equations using unconstrained string fields $V$ 
and $\Psi$. Thus this first-order formulation provides a systematic method to give
the equations of motion explicitly. We have clarified that 
the gauge transformation of $V$ and $\Psi$ are also obtained from
the gauge transformations in the first-order formulation, which are also written 
in a single transformation with the same form as that in the closed bosonic 
string field theory.

The most important and challenging problem in the WZW-like formulation 
is to construct an action including the R sector, which is not straightforward due to 
the picture number mismatch.\cite{Berkovits:2001im}
One promising construction proposed for the open superstring
was given by introducing an auxiliary R string field $\Xi$ with a constraint
$Q_J\Xi=\eta\Psi$ imposed after deriving the equations of motion.\cite{Michishita:2004by}
We have partially constructed such an action for the heterotic
string field theory with the constraint 
$Q_G\Xi=B_{-1/2}$,\cite{Kunitomo:2013mqa}\footnote{See also Appendix.} 
but not yet succeeded in completing it in a closed form.
Another interesting possibility is to find an action 
which leads to the first-order equations of motion (\ref{FOeom}).
Such a possibility was examined for the open superstring, 
proposing a new superstring field theory.\cite{Kroyter:2009rn}
In this formulation, the string fields have all the picture
numbers democratically, but it is unclear whether the theory
reproduces the correct physical spectrum or not due to the subtlety of 
the $\hat{Q}$-cohomology of such a space.\cite{Berkovits:2001us,Kroyter:2009rn}
For the equivalence between the first-order and the conventional formulations
of the heterotic string field theory, it is necessary to assume that 
only the single NS string field $B_0$ and the single R string field $B_{-1/2}$ 
are non-zero at the linearized level.
Another difficulty is that the democratic formulation
requires explicit insertion of the picture changing operators,
which seems to be more serious since there is no midpoint 
in the heterotic string. This difficulty may be overcome 
by using the formulation proposed recently.\cite{Erler:2013xta,Erler:2014eba}

It is also interesting that the the first-order 
equations of motion can be written in a single equation 
with the same form as the equation of motion of 
the closed bosonic string field theory.
This suggests that the equations of motion of the Type II superstring
field theory in the similar formulation can also be written
in a single equation with the same form.
It is worthwhile to consider such a possibility in detail, 
since it is difficult to construct an action of the type II theory
even for the NS-NS sector.\cite{Matsunaga}\cite{Erler:2014eba}
This is now under investigation.

\vspace{5mm}
\section*{Acknowledgments}

The author would like to thank Yuji Okawa for discussion.

\newpage
\vspace{2cm}
\appendix

\section{}\label{app}

The first-order formulation provides us a systematic and efficient method 
to obtain an explicit expression of the equations of motion (\ref{eom}).
We illustrate it with some new explicit results 
in this appendix. Some results on the action and 
the gauge transformations are also given.

From modified subsidiary equations (\ref{subsidiary2}), we have,
at the $(n+4)$-th order,
\begin{align}
S(1)_{-\frac{(n+2)}{2}}^{(n+4)}=\eta B_{-\frac{n}{2}}^{(n+4)} + Q_GB_{-\frac{(n+2)}{2}}^{(n+4)}
+&\sum_{m=1}^{n+1}\frac{\kappa^m}{m!}\mspd{(B_{-\frac{1}{2}}^{(1)})^m,
B_{-\frac{(n-m+2)}{2}}^{(n-m+4)}}
\nonumber\\
&-\frac{\kappa^{n+3}}{2(n+2)!}\mspd{\Psi,\epsi^{n+1},\mspd{(B_{-1/2}^{(1)})^2}}
=\ 0.\label{sube2}
\end{align}
Using  (\ref{linearized omega}) and (\ref{leading}), we can show that
these $S_{(n+2)/2}^{(n+4)}$ can be written as the sum of the $\eta$-exact terms and the terms
including either $Q_G\eta\Psi$ or $\eta G$:
\begin{align}
 \eta\Bigg( B_{-\frac{n}{2}}^{(n+4)}
&+\frac{\kappa^{n+3}}{(n+4)!}\mspd{\Psi,Q_G\Psi,\epsi^{n+2}}
+\frac{\kappa^{n+3}}{(n+4)!}(n+3)\mspd{\Psi,\epsi^{n+1},\mspd{\Psi,\eta\Psi}}\nonumber\\
&-\frac{\kappa^{n+3}}{(n+4)!}\sum_{m=0}^n\binom{n+3}{m}
\mspd{\Psi,\epsi^m,\mspd{\Psi,\epsi^{n-m+2}}}
\Bigg)\nonumber\\
&+\frac{\kappa^{n+3}}{(n+3)!}\mspd{\Psi,\epsi^{n+2},Q_G\eta\Psi}
-\frac{\kappa^{n+4}}{(n+4)!}\mspd{\Psi,Q_G\Psi,\epsi^{n+2},\eta G}
\nonumber\\
&-\frac{\kappa^{n+4}}{(n+4)!}\mspd{\Psi,\epsi^{n+2},\mspd{\Psi,\eta G}}
-\frac{\kappa^{n+4}}{(n+4)!}(n+3)\mspd{\mspd{\Psi,\eta\Psi},\Psi,\epsi^{n+1},\eta G}
\nonumber\\
&
-\frac{\kappa^{n+4}}{(n+4)!}(n+3)\mspd{\Psi,\epsi^{n+1},\mspd{\Psi,\eta\Psi,\eta G}}
\nonumber\\
&+\frac{\kappa^{n+4}}{(n+4)!}\sum_{m=0}^n\binom{n+3}{m}
\mspd{\mspd{\Psi,\epsi^{n-m+2}},\Psi,\epsi^m,\eta G}
\nonumber\\
&+\frac{\kappa^{n+4}}{(n+4)!}\sum_{m=0}^n\binom{n+3}{m}
\mspd{\Psi,\epsi^m,\mspd{\Psi,\epsi^{n-m+2},\eta G}}=0,\label{subeqs2}
\end{align}
where $\binom{n}{m}=\frac{n!}{(n-m)!m!}$.
Then we can solve them up to the equations of motion as
\begin{align}
 B_{-\frac{n}{2}}^{(n+4)} =&\ -\frac{\kappa^{n+3}}{(n+4)!}
\mspd{\Psi,Q_G\Psi,\epsi^{n+2}}
-\frac{\kappa^{n+3}}{(n+4)!}(n+3)\mspd{\Psi,\epsi^{n+1},
\mspd{\Psi,\eta\Psi}}\nonumber\\
&+\frac{\kappa^{n+3}}{(n+4)!}
\sum_{m=0}^n\binom{n+3}{m}\mspd{\Psi,\epsi^m,\mspd{\Psi,\epsi^{n-m+2}}}.
\label{nextleading}
\end{align}
Although the $\eta$-non-exact part of (\ref{subeqs2}) is much complicated
than (\ref{sube1-1}), it can be read from the consistency equations of 
the modified subsidiary equations (\ref{subsidiary2}),
\begin{align}
\eta S(1)_{-\frac{(n+2)}{2}}=&-Q_GS(1)_{-\frac{(n+4)}{2}}
-\sum_{l=1}^{n+1}\sum_{m=1}^{n-l+2}\frac{\kappa^m}{m!}
\mspd{\left(\tilde{B}^m\right)_{-\frac{(n-l+2)}{2}},S(1)_{-\frac{(l+2)}{2}}}\nonumber\\
&+\frac{\kappa^{n+2}}{(n+2)!}\mspd{\Psi,\epsi^{n+1},Q_GS(1)_{-2}}
+\frac{\kappa^{n+3}}{(n+2)!}\mspd{\Psi,\epsi^{n+1},\mspd{B_{-\frac{1}{2}},S(1)_{-\frac{3}{2}}}}
\nonumber\\
&-\sum_{m=1}^{n+3}\frac{\kappa^m}{m!}
\mspd{\left(\tilde{B}^m\right)_{-\frac{(n+3)}{2}},E_{-\frac{1}{2}}}
+\sum_{m=1}^3\frac{\kappa^{n+m+2}}{(n+2)!m!}\mspd{\Psi,\epsi^{n+1},
\mspd{(\tilde{B}^m)_{-\frac{3}{2}},E_{-\frac{1}{2}}}}\nonumber\\
&+\frac{\kappa^{n+5}}{(n+4)!}\mspd{\Psi,\epsi^{n+3},\mspd{B_{-\frac{1}{2}},E_{-\frac{1}{2}}}}
\nonumber\\
&-\frac{\kappa^{n+7}}{(n+2)! 4!}\mspd{\Psi,\epsi^{n+1},\mspd{\Psi,\epsi^3,
\mspd{B_{-\frac{1}{2}},E_{-\frac{1}{2}}}}}\nonumber\\
&-\sum_{m=1}^{n+2}\frac{\kappa^m}{m!}
\mspd{\left(\tilde{B}^m\right)_{-\frac{(n+2)}{2}},E_{-1}}
+\frac{\kappa^{n+2}}{(n+2)!}\mspd{\epsi^{n+2},E_{-1}}
\nonumber\\
&-\frac{\kappa^{n+3}}{(n+2)!}\mspd{\Psi,\epsi^{n+1},\eta G,E_{-1}}
+\frac{\kappa^{n+4}}{(n+4)!}\mspd{Q_G\Psi,\epsi^{n+3},E_{-1}}
\nonumber\\
&
-\frac{\kappa^{n+4}}{(n+4)!}(n+3)\mspd{\Psi,\epsi^{n+2},Q_G\eta\Psi,E_{-1}}\nonumber\\
&+\frac{\kappa^{n+4}}{(n+4)!}\sum_{m=2}^{n+3}
\binom{n+3}{m}\mspd{\mspd{\epsi^m},\Psi,\epsi^{n-m+3},E_{-1}}\nonumber\\
&+\frac{\kappa^{n+4}}{(n+4)!}\sum_{m=1}^{n+3}
\binom{n+3}{m}\mspd{\mspd{\Psi,\epsi^m},\epsi^{n-m+3},E_{-1}}\nonumber\\
&-\frac{\kappa^{n+4}}{(n+4)!}\sum_{m=1}^{n+3}
\binom{n+3}{m}\mspd{\Psi,\epsi^{n-m+3},\mspd{\epsi^m,E_{-1}}}\nonumber\\
&+\frac{\kappa^{n+4}}{(n+4)!}\sum_{m=0}^{n+2}
\binom{n+3}{m}\mspd{\epsi^{n-m+3},\mspd{\Psi,\epsi^m,E_{-1}}}\nonumber\\
&-\sum_{l=1}^{n+1}\sum_{m=1}^{n-l+2}\frac{\kappa^{m+l+2}}{(l+2)!m!}
\mspd{(\tilde{B}^m)_{-\frac{(n-l+2)}{2}},\mspd{\Psi,\epsi^{l+1},E_{-1}}}\nonumber\\
&-\frac{\kappa^{n+6}}{(n+2)! 4!}\mspd{\Psi,\epsi^{n+1},\mspd{Q_G\Psi,\epsi^3,E_{-1}}}\nonumber\\
&+\frac{3\kappa^{n+6}}{(n+2)! 4!}\mspd{\Psi,\epsi^{n+1},\mspd{\Psi,\epsi^2,Q_G\eta\Psi,E_{-1}}}
\nonumber\\
&-\frac{\kappa^{n+6}}{(n+2)! 4!}\sum_{m=2}^3\binom{3}{m}\mspd{\Psi,\epsi^{n+1},
\mspd{\mspd{\epsi^m},\Psi,\epsi^{3-m},E_{-1}}}\nonumber\\
&-\frac{\kappa^{n+6}}{(n+2)! 4!}\sum_{m=1}^3\binom{3}{m}
\mspd{\Psi,\epsi^{n+1},\mspd{\mspd{\Psi,\epsi^m},\epsi^{3-m},E_{-1}}}\nonumber\\
&+\frac{\kappa^{n+6}}{(n+2)! 4!}\sum_{m=1}^3\binom{3}{m}\mspd{\Psi,\epsi^{n+1},
\mspd{\Psi,\epsi^{3-m},\mspd{\epsi^m,E_{-1}}}}
\nonumber\\
&-\frac{\kappa^{n+6}}{(n+2)! 4!}\sum_{m=0}^2
\binom{3}{m}\mspd{\Psi,\epsi^{n+1},\mspd{\epsi^{3-m},\mspd{\Psi,\epsi^m,E_{-1}}}}
\nonumber\\
&+\sum_{m=1}^2\frac{\kappa^{n+m+2}}{(n+2)!m!}
\mspd{\Psi,\epsi^{n+1},\mspd{(\tilde{B}^m)_{-1},E_{-1}}}\nonumber\\
&+\frac{\kappa^{n+6}}{(n+2)!3!}\mspd{\Psi,\epsi^{n+1},\mspd{B_{-\frac{1}{2}},
\mspd{\Psi,\epsi^2,E_{-1}}}},
\label{conSub2}
\end{align}
which can be computed using (\ref{conSub}) and the consistency equations
of the equations of motion,
\begin{align}
Q_GE_{-\frac{1}{2}}=&0,\qquad 
Q_GE_{-1}+\kappa\mspd{B_{-\frac{1}{2}},E_{-\frac{1}{2}}}=0,\nonumber\\
\eta E_{-\frac{n}{2}}=&-Q_GS_{-\frac{(n+2)}{2}}
-(1-\delta_{n,1})\kappa\mspd{B_{-\frac{1}{2}},S_{-\frac{3}{2}}}\nonumber\\
&-\sum_{m=1}^{n}\frac{\kappa^m}{m!}
\mspd{\left(\tilde{B}^m\right)_{-\frac{n}{2}},E_{-1}}
-\sum_{m=1}^{n+1}\frac{\kappa^m}{m!}
\mspd{\left(\tilde{B}^m\right)_{-\frac{(n+1)}{2}},E_{-\frac{1}{2}}},
\qquad (n=1,2),
\label{conEOM}
\end{align} 
obtained from (\ref{conB}).
From these new consistency equations, we have
\begin{align}
\eta S(1)_{-\frac{(n+2)}{2}}^{(n+4)}=&
\eta\Bigg(\frac{\kappa^{n+3}}{(n+3)!}\mspd{\Psi,\epsi^{n+2},Q_G\eta\Psi}
-\frac{\kappa^{n+4}}{(n+4)!}\mspd{\Psi,Q_G\Psi,\epsi^{n+2},\eta G}
\nonumber\\
&-\frac{\kappa^{n+4}}{(n+4)!}\mspd{\Psi,\epsi^{n+2},\mspd{\Psi,\eta G}}
-\frac{\kappa^{n+4}}{(n+4)!}(n+3)\mspd{\mspd{\Psi,\eta\Psi},\Psi,\epsi^{n+1},\eta G}
\nonumber\\
&
-\frac{\kappa^{n+4}}{(n+4)!}(n+3)\mspd{\Psi,\epsi^{n+1},\mspd{\Psi,\eta\Psi,\eta G}}
\nonumber\\
&+\frac{\kappa^{n+4}}{(n+4)!}\sum_{m=0}^n\binom{n+3}{m}
\mspd{\mspd{\Psi,\epsi^{n-m+2}},\Psi,\epsi^m,\eta G}
\nonumber\\
&+\frac{\kappa^{n+4}}{(n+4)!}\sum_{m=0}^n\binom{n+3}{m}
\mspd{\Psi,\epsi^m,\mspd{\Psi,\epsi^{n-m+2},\eta G}}\Bigg),\label{subeqs2-2}
\end{align}
at the $(n+4)$-th order. The terms in the right hand side come from 
the $\eta$-non-exact part in (\ref{subeqs2}). 
Using these explicit results, the subsidiary equations (\ref{subsidiary2}) can 
further be modified as
\begin{align}
S(2)_{-\frac{(n+2)}{2}}=&S(1)_{-\frac{(n+2)}{2}}
-\frac{\kappa^{n+3}}{(n+3)!}\mspd{\Psi,\epsi^{n+2},E_{-\frac{1}{2}}}\nonumber\\
&+\frac{\kappa^{n+4}}{(n+4)!}\mspd{\Psi,Q_G\Psi,\epsi^{n+2},E_{-1}}
+\frac{\kappa^{n+4}}{(n+4)!}\mspd{\Psi,\epsi^{n+2},\mspd{\Psi,E_{-1}}}\nonumber\\
&+\frac{\kappa^{n+4}}{(n+4)!}(n+3)\mspd{\mspd{\Psi,\eta\Psi},\Psi,\epsi^{n+1},E_{-1}}
\nonumber\\
&+\frac{\kappa^{n+4}}{(n+4)!}(n+3)\mspd{\Psi,\epsi^{n+1},\mspd{\Psi,\eta\Psi,E_{-1}}}
\nonumber\\
&-\frac{\kappa^{n+4}}{(n+4)!}\sum_{m=0}^n\binom{n+3}{m}
\mspd{\mspd{\Psi,\epsi^{n-m+2}},\Psi,\epsi^m,E_{-1}}
\nonumber\\
&-\frac{\kappa^{n+4}}{(n+4)!}\sum_{m=0}^n\binom{n+3}{m}
\mspd{\Psi,\epsi^m,\mspd{\Psi,\epsi^{n-m+2},E_{-1}}}
=0,
\label{subsidiary3}
\end{align}
so as to be $\eta$-exact at the $(n+4)$-th order:
\begin{align}
S(2)_{-\frac{(n+2)}{2}}^{(n+4)}=&\eta\Bigg(B_{-\frac{n}{2}}^{(n+4)}
+\frac{\kappa^{n+3}}{(n+4)!}\mspd{\Psi,Q_G\Psi,\epsi^{n+2}}
+\frac{\kappa^{n+3}}{(n+4)!}(n+3)\mspd{\Psi,\epsi^{n+1},\mspd{\Psi,\eta\Psi}}\nonumber\\
&\hspace{1cm}
-\frac{\kappa^{n+3}}{(n+4)!}
\sum_{m=0}^n\binom{n+3}{m}\mspd{\Psi,\epsi^m,\mspd{\Psi,\epsi^{n-m+2}}}
\Bigg)=0,\label{sub2}
\end{align}
which can exactly be solved by (\ref{nextleading}).
Substituting this solutions into the new subsidiary equations (\ref{subsidiary3}),
they are again the sum of the $\eta$-exact part and the terms including either
$Q_G\eta\Psi$ or $\eta G$ at the next, $(n+6)$-th, order. 
We obtain, from the $\eta$-exact part,
\begin{align}
 B_{-\frac{n}{2}}^{(n+6)} =&
\frac{\kappa^{n+5}}{(n+6)!}\Bigg(
\mspd{\Psi,\qpsi^2,\epsi^{n+3}}\nonumber\\
&-\sum_{m=0}^n\frac{2n-m+10}{n+5}\binom{n+5}{m}
\mspd{\Psi,\epsi^m,\mspd{\Psi,Q_G\Psi,\epsi^{n-m+3}}}\nonumber\\
&+\sum_{m=n+1}^{n+3}\binom{n+4}{m+1}
\mspd{\Psi,\epsi^m,\mspd{\Psi,Q_G\Psi,\epsi^{n-m+3}}}\nonumber\\
&-\sum_{m=0}^{n+1}\binom{n+4}{m}
\mspd{\Psi,Q_G\Psi,\epsi^m,\mspd{\Psi,\epsi^{n-m+3}}}\nonumber\\
&+(2n+9)\mspd{\Psi,Q_G\Psi,\epsi^{n+2},\mspd{\Psi,\eta\Psi}}
\nonumber\\
&+\sum_{m=0}^n\sum_{l=0}^{n-m+1}\binom{n+5}{m}\binom{n-m+4}{l}
\mspd{\Psi,\epsi^m,\mspd{\Psi,\epsi^l,\mspd{\Psi,\epsi^{n-m-l+3}}}}\nonumber\\
&-\binom{n+5}{3}\mspd{\Psi,\epsi^{n+1},\mspd{\Psi,\mspd{\Psi,\epsi^2}}}
\nonumber\\
&-\sum_{m=0}^n\frac{(n-m+4)(2n-m+9)}{n+4}\binom{n+5}{m}
\mspd{\Psi,\epsi^m,\mspd{\Psi,\epsi^{n-m+2},\mspd{\Psi,\eta\Psi}}}\nonumber\\
&+(n+5)(n+3)\mspd{\Psi,\epsi^{n+1},\mspd{\Psi,\eta\Psi,\mspd{\Psi,\eta\Psi}}}
\nonumber\\
&+(n+5)\mspd{\Psi,\epsi^{n+2},\mspd{\Psi,\mspd{\Psi,\eta\Psi}}}
\nonumber\\
&+\sum_{m=0}^{n-1}\sum_{l=0}^{n-m-1}\frac{1}{2}\binom{n+5}{m+l+3}\binom{m+l+3}{m}
\mspd{\Psi,\epsi^m,\mspd{\Psi,\epsi^{l+2}},\mspd{\Psi,\epsi^{n-m-l+1}}}\nonumber\\
&-\sum_{m=0}^n(n+5)\binom{n+3}{m}
\mspd{\Psi,\epsi^m,\mspd{\Psi,\eta\Psi},\mspd{\Psi,\epsi^{n-m+2}}}\nonumber\\
&+(n+5)(n+3)\mspd{\Psi,\epsi^{n+1},\mspd{\Psi,\eta\Psi},\mspd{\Psi,\eta\Psi}}
\Bigg).
\end{align}
These results include the higher order corrections, $B_{-1/2}^{(7)}$ and 
$B_{-1}^{(8)}$, than those obtained previously.\cite{Kunitomo:2013mqa}

The string fields $B_{-n/2}$ built with small number of 
the string products $\mspd{\cdots}$ can be 
obtained to all orders,\cite{Okawa:2004ii,Kunitomo:2013mqa} 
starting from 
\begin{equation}
 B_{-\frac{1}{2}}^{[0]}=\eta\Psi,\qquad B_{-\frac{n}{2}}^{[0]}=0,\ (n\ge2),
\end{equation}
where the number in $[\cdot]$ at the superscript denotes the number of 
string products included. The subsidiary equations including the terms with 
a single string product can formally be written as
\begin{equation}
 \left(\eta B_{-\frac{k}{2}}^{[1]}\right)^{[1]}
+\left(Q_G B_{-\frac{(k+2)}{2}}^{[1]}\right)^{[1]}
+\frac{\kappa^{k+1}}{(k+2)!}\mspd{(B_{-\frac{1}{2}}^{[0]})^{k+2}}=0.\label{eq1}
\end{equation}
Note that $\eta B_{-n/2}^{[1]}$ and $Q_GB_{-n/2}^{[1]}$ are including
the terms with two or more string products coming through the equations
of motion or the fundamental relation (\ref{main}). 
The superscript $[1]$ at the outer parentheses denotes
the terms with a single string product among them.
Substituting the general form of terms with one string product,
\begin{equation}
 B_{-\frac{k}{2}}^{[1]}=\sum_{n=1}^\infty f_n^{(k)}\kappa^{2n+k-1}
\mspd{\Psi,\qpsi^{n-1},\epsi^{n+k}},
\end{equation} 
into (\ref{eq1}), we have
\begin{equation}
 f_1^{(k)}=\frac{1}{(k+2)!},\qquad
f_{n+1}^{(k)}+f_n^{(k+2)}=0,\label{rec1}
\end{equation}
for the numerical coefficient $f_n^{(k)}$. 
This recursion relation is easily solved by
\begin{equation}
 f_n^{(k)}=\frac{(-1)^{n+1}}{(2n+k)!}.
\end{equation}
Next, the terms built with two string products can generally be written as
\begin{equation}
B_{-\frac{k}{2}}^{[2]}=\sum_{n=1}^\infty\underset{(m,l)\ne(0,0)}{\sum_{m=0}^{n-2}\sum_{l=0}^{n+k}}
f_{n,m,l}^{(k)}\kappa^{2n+k-1}
\mspd{\Psi,\qpsi^{n-m-2},\epsi^{n-l+k},\mspd{\Psi,\qpsi^m,\epsi^l}}.\label{two products}
\end{equation}
Substituting these expressions into the subsidiary equations,
\begin{align}
 &\left(\eta B_{-\frac{k}{2}}^{[1]}\right)^{[2]}
+\left(Q_G B_{-\frac{(k+2)}{2}}^{[1]}\right)^{[2]}\nonumber\\
&\hspace{1cm}
+\left(\eta B_{-\frac{k}{2}}^{[2]}\right)^{[2]}
+\left(Q_G B_{-\frac{(k+2)}{2}}^{[2]}\right)^{[2]}
+\sum_{l=1}^{k+1}\frac{\kappa^l}{l!}\mspd{(B_{-\frac{1}{2}}^{[0]})^l,B_{-\frac{(k-l+2)}{2}}^{[1]}}=0,
\end{align}
we have
\begin{subequations} 
\begin{align}
&f_{n,0,l}^{(k)}-\binom{n+k+1}{l+1}f_{n-1}^{(k+2)}+\frac{1}{2}f_{n-1}^{(k)}\delta_{l,1}=0,
\qquad (n\ge2,\ 1\le l\le n+k),\\
&f_{n,n-2,l}^{(k)}+\binom{n+k+1}{l}f_{n-1}^{(k+2)}=0,\qquad
(n\ge2,\ 0\le l\le n+k),\\
&f_{n,m,0}^{(k)}-\binom{n-1}{m+1}f_n^{(k)}=0,\qquad
(n\ge3,\ 1\le m\le n-2),\\
&f_{n,m,n+k}^{(k)}+\binom{n-1}{m}f_n^{(k)}=0,\qquad
(n\ge 2,\ 0\le m\le n-2),\\
&f_{n+1,m,l}^{(k)}+f_{n,m,l}^{(k+2)}
+\binom{n-1}{m}\binom{n+k+2}{l}f_n^{(k+2)}=0,\nonumber\\
&\hspace{6cm}(n\ge2,\ 0\le m\le n-2,\ 0\le l\le n+k+1),\\
&f_{n+1,m,l-1}^{(k)}+f_{n,m-1,l}^{(k+2)}-\binom{n-1}{m}\binom{n+k+2}{l}f_n^{(k+2)}\nonumber\\
&\hspace{2cm}
+\binom{2n+k+2}{2m+1}f_n^{(k+2)}\delta_{l,m+1}
+\binom{2n+k+2}{2m+2}f_n^{(k+2)}\delta_{l,m+2}=0,\nonumber\\
&\hspace{6cm}
(n\ge 2,\ 1\le m\le n-1,\ 1\le l\le n+k+2),
\end{align}
\end{subequations}
for the coefficients $f_{n,m.l}^{(k)}$.
These recursion relations are solved by
\begin{equation}\label{coef2}
f_{n,m,l}^{(k)}=
\begin{cases}
\displaystyle
\sum_{i=0}^l\binom{n-1}{m+l+1-i}\binom{n+k+1}{i}
\frac{(-1)^{n+1}}{(2n+k)!}, & \textrm{for}\ (0\le l\le m+1), \\
\displaystyle
-\sum_{i=0}^m\binom{n-1}{i}\binom{n+k+1}{m+l+1-i}\frac{(-1)^{n+1}}{(2n+k)!},
 & \textrm{for}\  (m+2\le l\le n+k),
\end{cases}
\end{equation}
where the binomial coefficient $\binom{n}{m}$ was generalized as
\begin{equation}
 \binom{n}{m}=
\begin{cases}
\displaystyle \frac{n!}{(n-m)!m!} & \textrm{for}\ (0\le m\le n),\\
\ 0 & \textrm{for}\ (m<0,\ \textrm{or}\ \ m>n).
\end{cases}
\end{equation}
The results (\ref{two products}) with (\ref{coef2}) are including 
the correction $B_{-1/2}^{[2]}$ which was not obtained previously.
This enables us to write down the corrections 
to the action, built with two string products, as
\begin{align}
 S^{[2]}=&\sum_{n=2}^\infty\underset{(m,l)\ne(0,0)}{\sum_{m=0}^{n-2}\sum_{l=0}^{n+1}}
h_{n,m,l}\nonumber\\
&\hspace{1cm}
\times\kappa^{2n}\langle\eta\Psi,\mspd{\Psi,\qpsi^{n-m-2},\qxi^{n-l+1},
\mspd{\Psi,\qpsi^m,\qxi^l}}\rangle,\label{action}
\end{align}
where
\begin{equation}
 h_{n,m,l}=(n-l+2)f_{n,m,l}^{(2)}-\frac{1}{2}f_{n,m,l}^{(1)}
-\frac{1}{2}f_{n-m-1}^{(1)}f_{m+1}^{(1)}\delta_{l,m+2}.
\end{equation}
The part of the conventional equations of motion including two string products or less
are reproduced if we impose the constraint 
$Q_G\Xi=B_{-1/2}^{[0]}+B_{-1/2}^{[1]}+B_{-1/2}^{[2]}$ on the equations 
of motion derived from the action obtained previously\cite{Michishita:2004by,Kunitomo:2013mqa}
with the correction (\ref{action}).

The first-order formulation also gives a systematic method to obtain the gauge
transformations of $V$ and $\Psi$, while not as efficient as the one for 
the equations of motion. 
The gauge parameters at the $(n+4)$-th order are restricted as
\begin{align}
\sigma_{-n/2}^{(n+4)}=&-\frac{\kappa^{n+4}}{(n+4)!}\mspd{\Psi,Q_G\Psi,\epsi^{n+2},\eta\Lambda_1}\nonumber\\
&+\frac{\kappa^{n+4}}{(n+4)!}\sum_{m=0}^{n+1}\binom{n+3}{m}\mspd{\Psi,\epsi^m,\mspd{\Psi,\epsi^{n-m+2},\eta\Lambda_1}}
\nonumber\\
&+\frac{\kappa^{n+4}}{(n+4)!}\sum_{m=0}^n\binom{n+3}{m}\mspd{\mspd{\Psi,\epsi^{n-m+2}},\Psi,\epsi^m,\eta\Lambda_1}
\nonumber\\
&-\frac{n+3}{(n+4)!}\kappa^{n+4}\mspd{\mspd{\Psi,\eta\Psi},\Psi,\epsi^{n+1},\eta\Lambda_1}
+\frac{n+3}{(n+4)!}\kappa^{n+3}\mspd{\Psi,\epsi^{n+2},Q_G\Lambda_{1/2}}\nonumber\\
&-\frac{\kappa^{n+3}}{(n+4)!}\mspd{\Psi,\epsi^{n+2},\eta\Lambda_{3/2}},\qquad
n\ge0,\label{parameters at n+4}
\end{align}
from which we can compute the gauge transformations $B_\delta^{(4)}$ and $\delta^{(4)}\Psi$.
One can confirm the results reproduce the transformations given in Ref.~\citen{Kunitomo:2013mqa}.

\vfill
\pagebreak
%\vspace{2cm}

\end{document}